\documentclass[12pt]{article}
\usepackage{graphicx}
\usepackage{a4p}
\usepackage{amssymb}
\usepackage{cite}
\newcommand{\EPnum} {CERN-PH-EP-2004-035} 
\newcommand{\Date}  {12 July 2004}
\newcommand{\OPALColl}  {OPAL Collab.}

\newcommand{\NPhys}  {Nucl.~Phys.}
\newcommand{\NIM} {Nucl.~Instrum.\ Methods}

\newcommand{\CPC} {Comput. Phys. Commun.}
\newcommand {\lsp}      {{{\tilde{\chi}}^{0}}_{1}}
\newcommand {\nln}      {{{\tilde{\chi}}^{0}}_{2}}
\newcommand {\Gravitino} {\tilde{\mathrm{G}}}

\newcommand {\nngggbra}   {\nu\overline{\nu}\gamma\gamma(\gamma)}
\newcommand {\ra}         {\rightarrow}
\newcommand {\ee}         {\mathrm{e}^+ \mathrm{e}^-}

\newcommand {\Pep}     {\mathrm{e}^+}
\newcommand {\Pem}     {\mathrm{e}^-}
\newcommand {\Pgg}     {\gamma}
\newcommand {\Pgmp}    {\mu^+}
\newcommand {\Pgmm}    {\mu^-}
\newcommand {\Pgtp}    {\tau^+}
\newcommand {\Pgtm}    {\tau^-}

\newcommand {\PZz}     {\mathrm{Z}^0}

\newcommand{\epem}   {\Pep\Pem}
\newcommand{\gamgam} {\Pgg\Pgg}
\newcommand{\mumu}   {\Pgmp\Pgmm}
\newcommand{\tautau} {\Pgtp\Pgtm}

\newcommand{\eetomumu}     {\epem\to\mumu}
\newcommand{\eetotautau}   {\epem\to\tautau}

\newcommand{\BR}             {{\mathrm{BR}}}
\newcommand{\PX}             {{\mathrm X}}
\newcommand{\PY}             {{\mathrm{Y}}}

\newcommand{\eetonngggbra}   {\epem \to \nngggbra}

\newcommand{\eetoXX}     {\epem \to \PX\PX}
\newcommand{\XtoYg}       {\PX \to \PY\gamma} 

\newcommand{\sigXX}       {\sigma(\epem \to \PX\PX)}
 
\newcommand{\sigbrXX}    {\sigma(\eetoXX)\cdot\BR^2(\XtoYg)} 
\newcommand{\eetoXXs}   {\epem \to \nln \nln}

\newcommand{\XtoYgs}    {\nln \to \lsp\gamma}

\newcommand{\mx}         {M_{\PX}}
\newcommand{\my}         {M_{\PY}}
\newcommand{\myzero}        {\my\approx 0}
\newcommand{\betax}      {\beta_{\mathrm X}}
\newcommand{\mxmax}      {\mx^{\rm max}}

\newcommand{\acosthe}    {|{\rm cos}\theta|}
\newcommand{\roots} {\sqrt{s}}

\begin{document}
\begin{titlepage}
\begin{center}{\large   EUROPEAN ORGANIZATION FOR NUCLEAR RESEARCH
}\end{center}\bigskip
\begin{flushright}
  \EPnum\\
  \Date \\
\end{flushright}

\begin{center}{\Large\bf \boldmath 
  Multi-Photon Events with Large Missing Energy \\
  in $\epem$ Collisions at $\roots = 192-209$ GeV
}\end{center}\bigskip\bigskip
\begin{center}{\LARGE The OPAL Collaboration
}\end{center}\bigskip \bigskip

\begin{abstract} 
Events with a final state consisting of two or more photons and large missing transverse 
energy have been observed in $\mathrm{e^+e^-}$ collisions at centre-of-mass energies 
in the range $192-209$ GeV using the OPAL detector at LEP. 
Cross-section measurements are performed within the kinematic 
acceptance of the selection and compared with the expectations from
the Standard Model process $\eetonngggbra$. No evidence 
for new physics contributions to this final state is observed. 
Upper limits on $\sigbrXX$ are derived for the case of stable and invisible 
$\PY$. In the case of massive $\PY$ the combined limits obtained from all 
the data range from 10~fb to 60~fb, while for the special case of massless $\PY$
the range is 20 fb to 40 fb. The limits apply to pair production of excited neutrinos 
($\PX = \nu^*, \PY = \nu$), to neutralino production ($\PX=\nln, \PY=\lsp$) 
and to supersymmetric models in which $\PX = \lsp$ and $\PY=\Gravitino$ is a
light gravitino. 
\end{abstract}





\bigskip\bigskip
\begin{center}{\large
To be submitted to 
Physics Letters {B}
}\end{center}
\end{titlepage}
\begin{center}{\Large        The OPAL Collaboration
}\end{center}\bigskip
\begin{center}{
G.\thinspace Abbiendi$^{  2}$,
C.\thinspace Ainsley$^{  5}$,
P.F.\thinspace {\AA}kesson$^{  3,  y}$,
G.\thinspace Alexander$^{ 22}$,
J.\thinspace Allison$^{ 16}$,
P.\thinspace Amaral$^{  9}$, 
G.\thinspace Anagnostou$^{  1}$,
K.J.\thinspace Anderson$^{  9}$,
S.\thinspace Arcelli$^{  2}$,
S.\thinspace Asai$^{ 23}$
% ...
}\end{center}\bigskip
\bigskip
$^{  1}$School of Physics and Astronomy, University of Birmingham,
Birmingham B15 2TT, UK
\newline
$^{  2}$Dipartimento di Fisica dell' Universit\`a di Bologna and INFN,
I-40126 Bologna, Italy
\newline
$^{  3}$Physikalisches Institut, Universit\"at Bonn,
D-53115 Bonn, Germany
\newline
% ...
$^{  5}$Cavendish Laboratory, Cambridge CB3 0HE, UK
\newline
% ...
$^{  9}$Enrico Fermi Institute and Department of Physics,
University of Chicago, Chicago IL 60637, USA
\newline
% ...
$^{ 16}$Department of Physics, Schuster Laboratory, The University,
Manchester M13 9PL, UK
\newline
% ...
$^{ 22}$Department of Physics and Astronomy, Tel Aviv University,
Tel Aviv 69978, Israel
\newline
$^{ 23}$International Centre for Elementary Particle Physics and
Department of Physics, University of Tokyo, Tokyo 113-0033, and
Kobe University, Kobe 657-8501, Japan
\newline
% ...
$^{  e}$ and Department of Experimental Physics, University of Debrecen, 
Hungary
\newline
% ...
$^{  y}$ now at CERN
\newline
% ...
%
\clearpage\newpage
\section{ Introduction }
\label{sec:intro}
 
We describe measurements and searches performed using a data sample of photonic events 
with large missing transverse energy collected with the OPAL detector in 1999 and 2000, 
the final
two years of LEP operation. The events result from $\epem$ collisions in the 
centre-of-mass energy range of about $192-209$~GeV with a combined integrated 
luminosity of 426.5~pb$^{-1}$. When deriving cross-section limits on new physics 
processes, these data are combined with previously published data\cite{OPALSP189}
taken at 189~GeV and corresponding to 177.3~pb$^{-1}$. The present paper builds on past 
publications based on data samples collected at lower centre-of-mass 
energies\cite{OPALSP189,OPALSP183,OPALSP172}. The new data samples, taken at the highest 
energies achieved by LEP, provide discovery potential in a new kinematic regime with 
a large increase in integrated luminosity. 
Similar searches have been made by the other 
LEP collaborations~\cite{LEP2AP}. 

The analysis presented here is designed to select events with 
two photons and significant missing transverse energy in the final state, indicating 
the presence of at least one neutrino-like invisible particle which interacts only 
weakly with matter. The event selection for this search topology is identical to that 
used in our most recent publication\cite{OPALSP189}. Within the Standard Model, such 
events are expected from the $\eetonngggbra$ process. The selection is designed to 
retain acceptance for events with an additional photon, provided that the system formed 
by the three photons is consistent with the presence of significant missing transverse 
energy.  

This final-state topology is also sensitive to several new physics 
scenarios. In the context of the search for new physics, the emphasis in this
publication is on general searches applicable to a broad class of models.
To this end, a generic classification is used: $\eetoXX$ where $\PX$ is
neutral and can decay radiatively ($\XtoYg$) and $\PY$ is stable and only weakly 
interacting. The limits presented for this generic process are applicable to a
variety of physics searches. For the general case of massive $\PX$ and $\PY$ this 
includes conventional supersymmetric processes $(\PX = \nln, \PY = \lsp)$. 
There is particularly good sensitivity for the special case of $\myzero$.
This is applicable both to the production of excited neutrinos $(\PX = \nu^*, \PY = \nu)$ 
and to supersymmetric models in which the lightest supersymmetric particle (LSP) is 
a light gravitino and $\lsp$ is the next-to-lightest supersymmetric particle 
(NLSP) which decays to a gravitino and a photon ($\PX=\lsp, \PY=\Gravitino$). 
In the latter case, we also set limits on an example light-gravitino 
model\cite{chang}. The neutralino lifetime in such models is a free parameter. In this 
paper we address only the case of promptly decaying~$\PX$. 

This search topology also has sensitivity to the production of two particles, one 
invisible, or with an invisible decay mode, and the other decaying into two photons. 
Such events might arise from the production of a Higgs-like scalar particle, 
$\rm S^0$ : $\rm e^+e^-\rightarrow Z^0S^0$, followed by S$^0$$\rightarrow \gamma\gamma$, 
$\rm Z^0\rightarrow \nu\overline{\nu}$. The results of an OPAL search for this process, 
including the hadronic and leptonic $\rm Z^0$ decays, have been separately 
reported~\cite{OPAL_Hgg209}. Finally, this search topology can also probe 
WW$\gamma \gamma$ quartic couplings in the 
$\ee \ra \nu_{\mathrm{e}} \overline{\nu_{\mathrm{e}}} \gamma \gamma$ 
process. The OPAL quartic gauge coupling measurements are described in~\cite{OPALQGC}.

This paper first describes the OPAL detector and the Monte Carlo samples 
used. A brief summary of the event selection will then be given, followed by 
cross-section measurements and comparisons with Standard Model expectations. 
The new physics search results will then be discussed. 

\section{OPAL Detector and Monte Carlo Samples}
\label{sec:detector}

The OPAL detector, which is described in detail in~\cite{OPAL-detector},
contained a silicon micro-vertex detector surrounded by a pressurized
central tracking system operating inside a solenoid with a magnetic field 
of 0.435 T. The barrel and endcap regions of the detector
were instrumented with scintillation counters, presamplers and a
lead-glass electromagnetic calorimeter (ECAL). The magnet return yoke was
instrumented for hadron calorimetry and was surrounded by muon chambers.
Electromagnetic calorimeters close to the beam axis measured luminosity and 
completed the acceptance.

The measurements presented here are based mainly on the observation of 
clusters of energy deposited in the lead-glass electromagnetic calorimeter. 
This consisted of an array of 9,440 lead-glass blocks in the barrel region, 
$|\cos{\theta}| < 0.82$, with a quasi-pointing geometry and two 
endcap arrays, each of 1,132 lead-glass blocks, covering the polar 
angle\footnote{The OPAL right-handed coordinate system is defined such that the origin 
is at the centre of the detector and the $z$ axis points along the direction of the 
$e^-$ beam.
The polar angle $\theta$ is defined with respect to the $e^-$ beam direction and 
$\phi$ is the azimuthal angle measured from the $+x$ axis.} 
range, $0.81<|\cos{\theta}|<0.984$. Hermetic electromagnetic 
calorimeter coverage was achieved beyond the end of the ECAL down to 33 mrad in
polar angle with the use of the gamma-catcher calorimeter, the forward 
calorimeter and the silicon-tungsten calorimeter. 

Scintillators in the barrel and endcap regions were used to reject backgrounds 
from cosmic-ray interactions by providing time measurements for the large 
fraction ($\approx$ 80\%) of photons which converted in the material in front of 
the ECAL. The barrel time-of-flight (TOF) scintillator bars were located outside 
the solenoid in front of the barrel ECAL and matched its geometrical acceptance 
$|\cos{\theta}| < 0.82$. Tile endcap (TE) scintillator arrays
were located in front of the endcap ECAL at $0.81<\acosthe <0.955$. 
Additional scintillating-tile arrays, referred to as the MIP plug,
were located at more forward angles. In the region from 125 to 200 mrad
these detectors were used to provide redundancy in the rejection of events
with significant electromagnetic activity in the forward region.

The integrated luminosities of the data samples are determined to better than 
1\% from small-angle Bhabha scattering events in the silicon-tungsten calorimeter. 
Triggers
based on electromagnetic energy deposits in either the 
barrel or endcap electromagnetic calorimeters lead to full trigger efficiency 
for photonic events passing the event selection criteria used in this analysis.
 
The NUNUGPV98~\cite{NUNUGPV98} and KK2f~\cite{KK2f} Monte Carlo generators were 
used to simulate the Standard Model signal process,  $\eetonngggbra$.
For other expected Standard Model processes, a number of different generators were used: 
RADCOR~\cite{RADCOR} for $\epem \to \gamma \gamma (\gamma)$; BHWIDE~\cite{BHWIDE} and 
TEEGG~\cite{TEEGG} for $\epem \to \epem (\gamma)$; KORALW~\cite{KORALW} 
using \texttt{grc4f}~\cite{GRC4F} matrix elements for 
$\ee \to \nu \bar{\nu} \ell^+ \ell^- (\gamma)$ and
$\ee \to \nu \bar{\nu} \mathrm{q} \bar{\mathrm{q}} (\gamma)$, 
and KORALZ\cite{KORALZ} for  $\eetomumu(\gamma)$ and $\eetotautau(\gamma)$.
The BDK program\cite{BDK} was used for $\ee \to \ee \ell^+ \ell^-$, except for
$\ee \to \ee\ee$ which was generated using the Vermaseren program\cite{VERMASEREN}.
The expected contribution from each of these Standard Model processes was
evaluated using a total equivalent integrated luminosity at least five times 
larger than the integrated luminosity of the data sample.

To simulate possible new physics processes of the type $\eetoXX$ where $\PX$ decays 
to $\PY\gamma$ and $\PY$ escapes detection, a modified version of the SUSYGEN~\cite{SUSYGEN} 
Monte Carlo generator was used to produce neutralino pair events of the type $\eetoXXs$, 
$\XtoYgs$, with isotropic angular distributions for the production and decay of $\nln$ and 
including the effects of initial-state radiation. For $\roots =$ 206 GeV, Monte Carlo events were 
generated at 49 points in the kinematically accessible region of the ($\mx$, $\my$) plane. 
Monte Carlo events at 42 points in ($\mx$, $\my$) with  $\roots =$ 189 GeV were generated for our
previous publication\cite{OPALSP189}. Using these two samples, the selection efficiency was 
determined for 
each generated point and then parametrized as a function of ($\mx$, $\my$) and centre-of-mass energy. 
The efficiency varies slowly with energy and for energies above 206 GeV, the 206~GeV values 
were used. All Monte Carlo samples described above were processed through the full OPAL detector 
simulation~\cite{GOPAL}.

\section{Event Selection }
\label{sec:selection}

A detailed description of the event selection is given in our previous
publications\cite{OPALSP189,OPALSP183}. In brief, photons are identified
as energy deposits in the electromagnetic calorimeter. Events are required
to have no other significant activity, except for the possibility of 
additional photons. Information from the tracking chambers is used to reject 
electromagnetic clusters associated with prompt charged tracks while retaining 
sensitivity for photons which converted in the material between the interaction 
point and the calorimeter. Timing information is used to reject backgrounds 
from cosmic-ray events. Events with activity beyond the acceptance of the ECAL
are vetoed using information from the gamma catcher, the forward
calorimeter, the silicon-tungsten calorimeter and the MIP plug.The kinematic 
acceptance of the selection is defined by requiring:

\begin{itemize}
\item at least two photons, each with $x_{\gamma} > 0.05 $
and $15^{\circ}<\theta<165^{\circ}$, or one photon 
with $E_{\gamma} > 1.75$ GeV  and $\acosthe < 0.8$ and a second photon
with $E_{\gamma} > 1.75$ GeV and $15^{\circ}<\theta<165^{\circ}$; here 
$E_{\gamma}$ is the photon energy, $\theta$ is the photon polar angle and
$x_{\gamma}$ is the photon scaled energy $E_{\gamma}/E_{\mathrm{beam}}$

\item that the two-photon system consisting of the two highest-energy photons 
have momentum transverse to the beamline ($p_T^{\gamma\gamma}$) satifying 
$p_T^{\gamma\gamma}/E_{\rm beam} > 0.05$
\end{itemize}

The selection is designed to retain acceptance for events with additional photons
in which the resulting photonic system is still consistent with the presence of 
significant missing energy. This reduces the sensitivity of the measurement to 
the modelling of higher-order contributions. 

\section{Selection Results}
\label{sec:results}

The data described in this paper were taken during the final two years of 
LEP operation, at centre-of-mass energies between 192 and 209 GeV. For the purposes
of this publication the data have been binned into six samples with mean 
centre-of-mass energies of approximately 192, 196, 200, 202, 205 and 207 GeV.
The energy ranges and luminosity breakdown are summarized in Table~\ref{tab:g2_xsec_new}. 
Applied to the entire sample, the selection yields a total of 
$54$ events, in good agreement with the KK2f prediction of $57.2 \pm 1.3$ events 
for the Standard Model $\eetonngggbra$ contribution. The expected contribution 
from other Standard Model processes and from cosmic ray and beam-related 
backgrounds is $1.2\pm 0.3$ events, dominated by contributions from low-angle radiative 
Bhabha events and radiative four-fermion final states. 
The selection results are included in Table~\ref{tab:g2_xsec_new}. The selection 
efficiency for $\eetonngggbra$ 
events within the kinematic acceptance of the selection is ($65.7{\pm 1.5}$)\%, 
independent of energy. The cross-section within the kinematic acceptance of the selection
is also shown in Table~\ref{tab:g2_xsec_new} as are the corresponding predictions obtained
using the KK2f Monte Carlo generator. 
The predictions of the NUNUGPV98 Monte Carlo generator were also examined 
and agreed well with those of KK2f. Small differences are accounted for in the 
systematic uncertainties.

The dominant sources of systematic uncertainties arise from modelling of the event selection 
efficiency, especially the simulation of the detector material and consequent photon 
conversion probabilities. The effects of these uncertainties and of uncertainties on
the efficiency of timing cuts used to suppress cosmic-ray events are calculated accounting for 
different event topologies (both photons in the barrel region, both in the endcap, or one in each). 
This total uncertainty is 1.7\%. Other sources arise from uncertainties on the 
integrated luminosity measurement (0.5\%), on  detector occupancy estimates (1\%) obtained 
from the analysis of randomly triggered events, on comparisons of different Monte Carlo event 
generators for the process $\eetonngggbra$(1\%). The total systematic uncertainty common to 
each energy bin is 2.3\%. In individual energy bins, Monte Carlo statistics account for an
additional systematic uncertainty of $0.9-1.4$\%.

\begin{figure}[b]
\begin{center}
\includegraphics[width=0.95\textwidth]{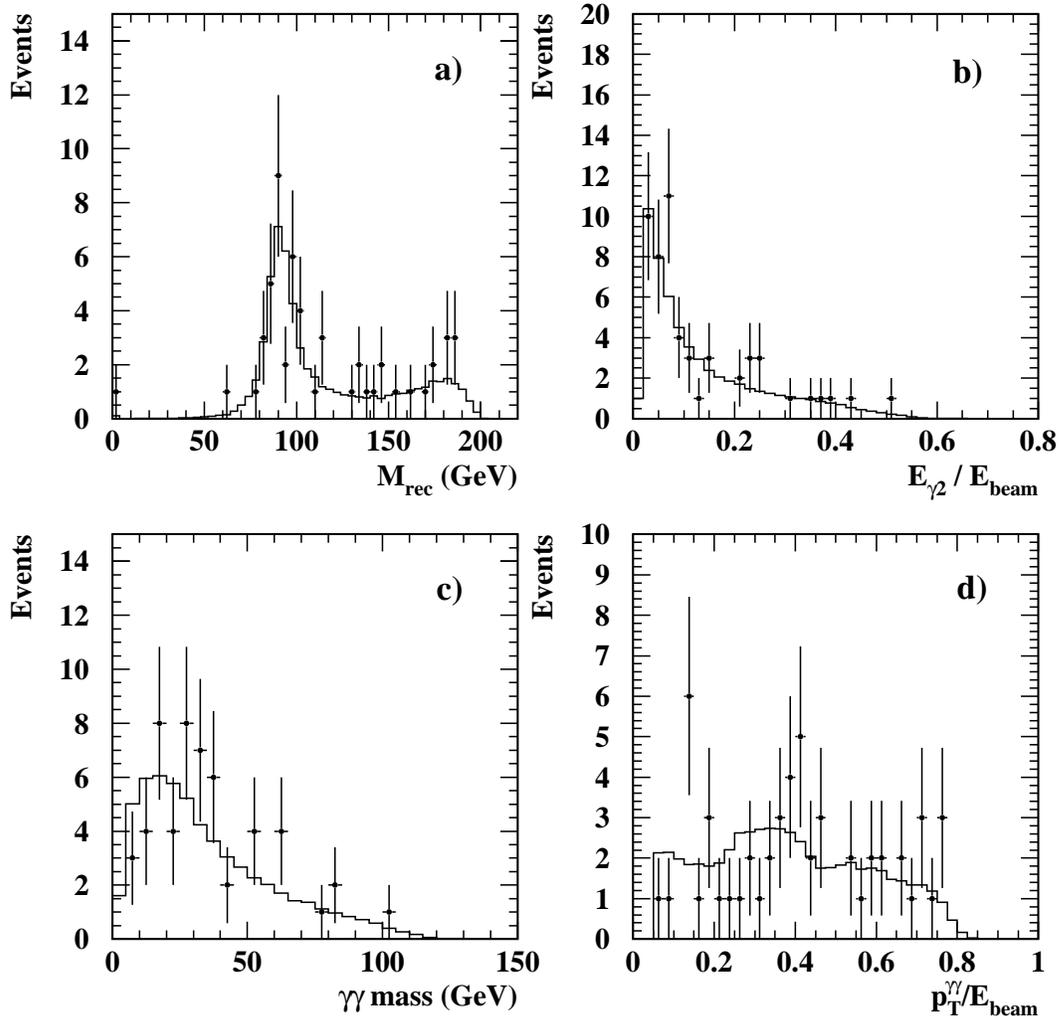}
\end{center}
\caption{Kinematic quantities of selected multi-photon events. 
Shown are 
a) the recoil-mass distribution b) the distribution of the scaled energy
of the second photon c) the distribution of the invariant mass
of the $\gamgam$ system and d) the scaled transverse momentum distribution 
for the $\gamgam$ system. The data points with error bars represent the 
selected OPAL data events. In each case the histogram shows the expected 
contribution from $\eetonngggbra$ events, from KK2f, normalized to the 
integrated luminosity of the data. The expected background from other sources
($1.2\pm{0.3}$ events) is not shown. 
}
\label{f:g2_kine_all}
\end{figure}

The kinematic properties of the selected events, summed over all energies, 
are displayed in Figure~\ref{f:g2_kine_all} where they are compared with 
the predicted distributions for $\eetonngggbra$ obtained using the KK2f 
generator normalized to the integrated luminosity of the data. Plot
(a) shows the recoil mass distribution of the selected events 
(for the two most energetic photons in the case of events with three or more photons). 
The distribution is  peaked near the mass of the 
$\PZz$ as is expected for contributions from $\eetonngggbra$.  The resolution 
of the recoil mass is typically $4-6$ GeV for $M_{\rm recoil}\approx M_{\rm Z}$.  
Events with a negative recoil-mass squared are plotted in the zero bin of the 
distribution.
Plot (b) shows the distribution of the scaled energy of the second most 
energetic photon. Plot (c) shows the $\gamma\gamma$ invariant-mass distribution 
for which the mass resolution is typically $1-2$ GeV. Plot (d) shows the
distribution in scaled transverse momentum of the selected two-photon system. 

There are 3 selected events having a third photon with deposited energy above 
300 MeV and within the polar-angle acceptance of the selection. The corresponding
expectation  from KK2f is $3.36 \pm 0.08$ events. 

\section{Data Interpretation}
\label{sec:interpret}

The results of this selection are used to test the Standard Model and to search 
for new physics contributions. 
%
%
In the absence of an excess of events beyond the Standard Model expectation, 
we set 95\% CL upper limits on the
quantity $\sigbrXX$ for the general case of massive $\PX$ and $\PY$, 
and separately for the special case of $\myzero$. Efficiencies were 
evaluated under 
the assumption that $\PX$ decays promptly. Monte Carlo samples were 
generated for a variety of mass points in the kinematically accessible region of 
the $(\mx,\my)$ plane. To set limits for arbitrary $\mx$ and $\my$, the 
efficiency over the entire $(\mx,\my)$ plane was parameterized using the 
efficiencies calculated at the generated mass points. For $\mx$ values 
below $M_{\rm Z}$/2, search results based on LEP1 data have been previously
reported~\cite{LEP1XX}. In this low-mass region, events with radiative return 
to the $\PZz$ followed by $\PZz\rightarrow\PX\PX$ would yield very different kinematics
than those used here to generate the signal Monte Carlo samples. For this reason,
the search is restricted to the mass region $\mx > M_{\mathrm Z}/2$.

\boldmath
\subsection{Search for $\eetoXX$, $\XtoYg$ ; General case: $\my\ge 0$}
\unboldmath
\label{sec:g2_results_allmy}

The searches for $\eetoXX$, $\XtoYg$, both for the general case discussed here 
and the special case of $\myzero$ discussed in section 5.2, use the methods described 
in our previous publications\cite{OPALSP189,OPALSP183}. Selected events are 
classified as consistent with a given value of $\mx$ and $\my$ if the energy of 
each of the photons falls within the region kinematically accessible to photons 
from the process $\eetoXX$, $\XtoYg$, including resolution effects. Selection
efficiencies at some of the generated grid points for the $\eetoXX$, $\XtoYg$ 
$\sqrt{s}=206$~GeV Monte Carlo events are shown in Table~\ref{tab:g2_eff_206}. 
These values include the efficiency of the kinematic consistency requirement which 
is higher than 95\% at each generated point in the region of the $(\mx,\my)$ plane. 
For $\mx-\my$ values lower than 5 GeV the efficiency begins to fall off rapidly
and is thus difficult to model accurately. For this reason, we place limits only 
in the region of the $(\mx, \my)$ plane satisfying $\mx-\my \ge 5$ GeV. Efficiencies 
at lower centre-of-mass energies are obtained from an interpolation between these 
efficiencies and the equivalent efficiencies at 189 GeV, which are given in our 
previous publication~\cite{OPALSP189}. For data taken at centre-of-mass energies
above 206 GeV, the 206 GeV efficiencies are used.

\begin{figure}[b]
\begin{center}
\includegraphics[width=0.95\textwidth]{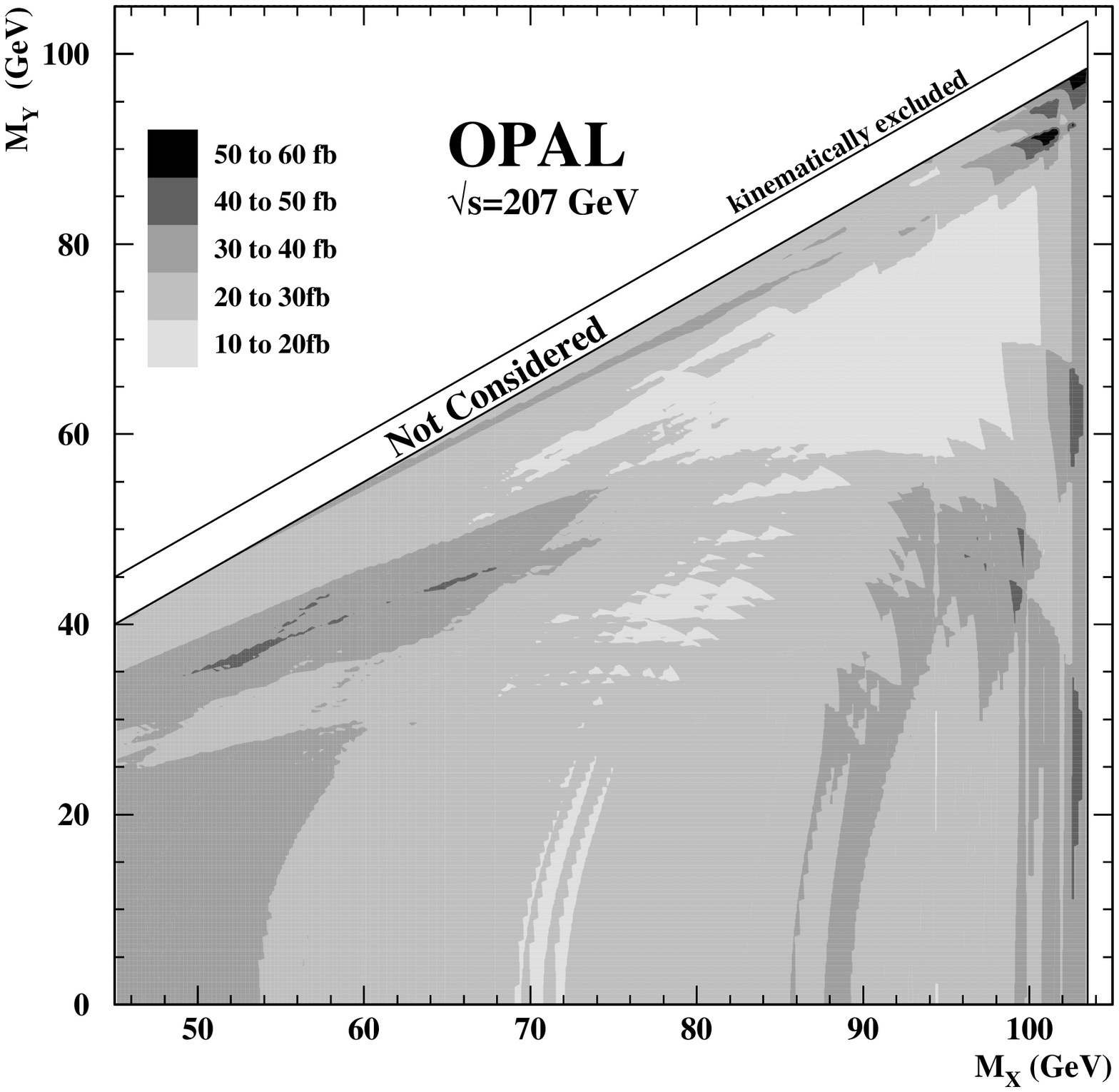}
\end{center}
\caption{
The shaded areas show 95\% CL upper limits
on the quantity $\sigbrXX$ at $\roots = 207$ GeV obtained 
from all OPAL data with $\roots\ge$ 189 GeV, under the 
assumption that 
the cross-section scales as $\betax/s$. No limit is set 
for mass-difference values $\mx-\my < 5$ GeV, defined by 
the lower line above the shaded regions. The upper line is 
for $\mx=\my$.
}
\label{mxmy_207_combined}
\end{figure}

Events from $\eetonngggbra$ are typically characterized by a high-energy photon 
from the radiative return to the $\rm Z^0$ and a second lower energy photon. 
The kinematic consistency requirement is such that the two photons must have 
energies within the same (kinematically accessible) region. Thus, as $\mx$ and $\my$ 
increase, the allowed range of energy for the photons narrows, and fewer 
$\nngggbra$ events will be accepted. For the 54 selected events, the distribution
of the number of events consistent with a given mass point ($\mx$,$\my$) is 
consistent with the expectation from $\eetonngggbra$ Monte Carlo, over the full
($\mx$,$\my$) plane.
Upper limits are placed on $\sigbrXX$ accounting for the
number of selected events and the expected number of background events from the process
$\eetonngggbra$. Other backgrounds are not subtracted. For each of the energy bins, 
Table~\ref{tab:limits} shows the maximum and minimum limits obtained in the 
region of the ($\mx, \my)$ plane described above. Figure~\ref{mxmy_207_combined} shows 
the 95\% CL lower limits on $\sigbrXX$ at $\sqrt{s}=207$ GeV, obtained from all OPAL data with 
$\sqrt{s}\ge 189$ GeV, under the assumption that $\sigXX$ scales with centre-of-mass 
energy as $\betax/s$. These limits range from $10-60$ fb.

Systematic uncertainties arise from the sources described in section 4. However
there are additional contributions due to limited Monte Carlo statistics at each 
of the generated ($\mx,\my$) points and from uncertainties on the efficiency 
parameterization across the ($\mx,\my$) plane and as a function of 
energy. The combined relative uncertainty on the efficiency varies from about 3\% to 6\% 
across the plane (for $\mx - \my > 5$ GeV). The uncertainty on the expected SM 
background contribution is 2.6\%. In calculating the limits, systematic uncertainties are 
accounted for in the manner advocated in reference\cite{systerr}. This also applies to the 
limits for the $\myzero$ case, presented in the next section.

\boldmath
\subsection{Search for $\eetoXX$, $\XtoYg$ ; Special case: $\myzero$}
\unboldmath
\label{sec:g2_results_my0}

For the special case of $\myzero$ the applied kinematic consistency requirements differ 
from those used for the general case. One can calculate\cite{gravitinos2} the 
maximum mass, $\mxmax$, which is consistent with the measured three-momenta of 
the two photons, assuming a massless~$\PY$. A cut on $\mxmax$  provides further 
suppression of the $\nngggbra$ background while retaining high efficiency for 
the  signal hypothesis. This is discussed in more detail in 
reference~\cite{OPALSP172}. To allow for resolution effects, we require that the 
maximum kinematically allowed mass be greater than $\mx-5$ GeV. This has 
better than 96\% relative efficiency for signal at all values of $\mx$ 
while suppressing much of the remaining $\nngggbra$ background. 

\begin{figure}[b]
\begin{center}
\includegraphics[width=0.95\textwidth]{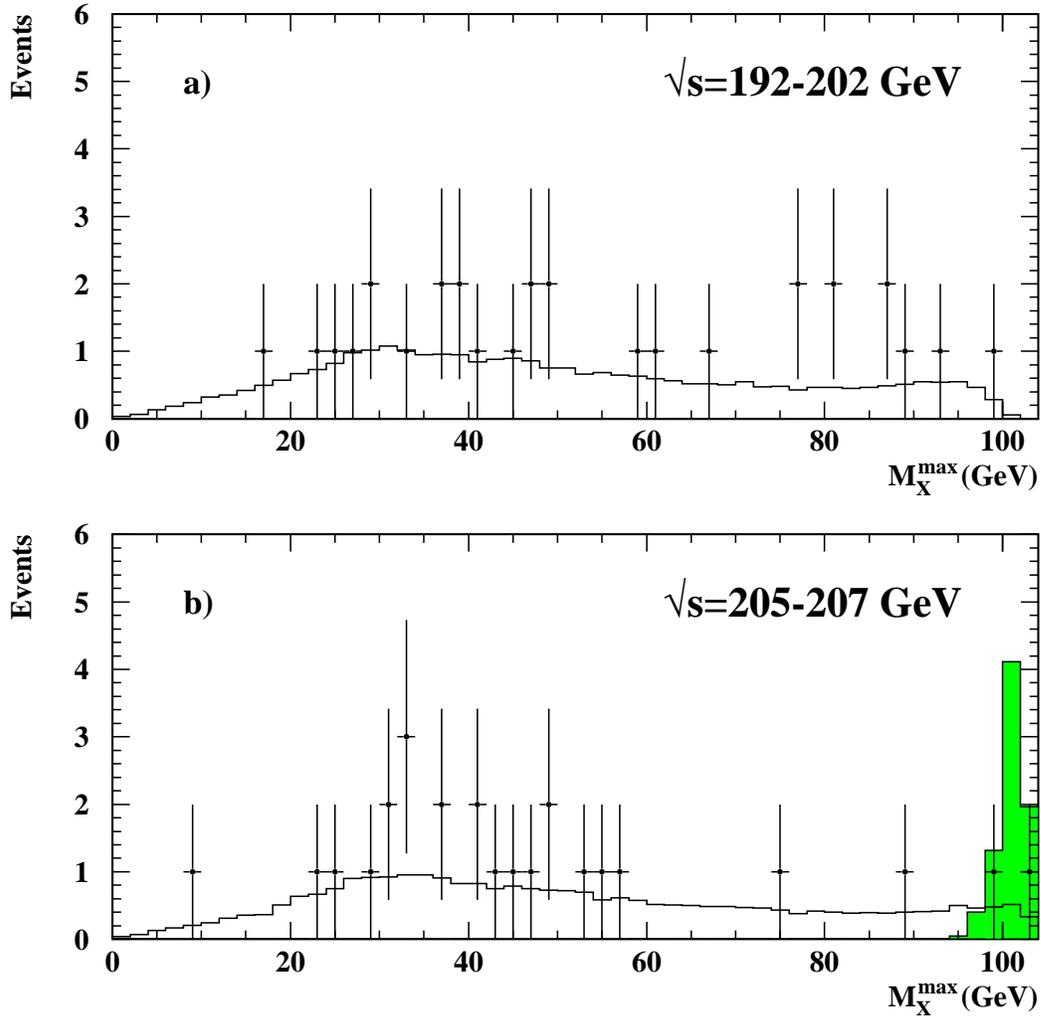}
\end{center}
\caption{The calculated value of $\mxmax$ for events selected
from a) the $192-202$ GeV data sample and b) the $205-207$ GeV sample. 
In each case the data points show the OPAL data and the unshaded histogram shows  
the expected distribution from the Standard Model process $\eetonngggbra$, 
evaluated using KK2f and normalized to the integrated luminosity of the 
data sample. In b) the shaded histogram shows the expected distribution for the signal 
process $\eetoXX$, $\XtoYg$ for $\mx = 100$ GeV with arbitrary 
production cross-section.}
\label{g2_mxmax_206}
\end{figure}

The $\mxmax$ distributions for all selected events, divided into the $192-202$~GeV 
and $205-207$~GeV data samples, are shown in Figure~\ref{g2_mxmax_206}.
In each case, the points with error bars show the OPAL data while the unshaded
histogram shows the expected contribution from the $\eetonngggbra$, from KK2f
Monte Carlo, normalized to the luminosity of the data.  
Shown as a shaded histgram in the $205-207$ GeV plot is the expected distribution from
signal Monte Carlo events generated with $\mx = 100$ GeV (with arbitrary normalization).
For this  $\myzero$ case, the signal reconstruction efficiencies 
calculated from Monte Carlo events generated at $\sqrt{s}=206$ GeV are shown 
in Table~\ref{tab:g2_eff_my0_206} after application of the event selection 
criteria and then after the cut on $\mxmax$. Also shown in 
Table~\ref{tab:g2_eff_my0_206} are the numbers of events selected from the 
$205-207$ GeV data sample which are consistent with each value of $\mx$ as well 
as the expected number of $\eetonngggbra$ events. The number of selected 
events (from the $205-207$ GeV sample) consistent with a given value of 
$\mx$ varies from 10, for $\mx\ge$ 45 GeV, to 2 at the kinematic limit. 
The expected number of events decreases from $14.9 \pm 0.4$ at $\mx\ge 45$ GeV 
to $1.28 \pm 0.08$ consistent with $\mx \ge 102.5$~GeV.

\begin{figure}[b]
\begin{center}
\includegraphics[width=0.95\textwidth]{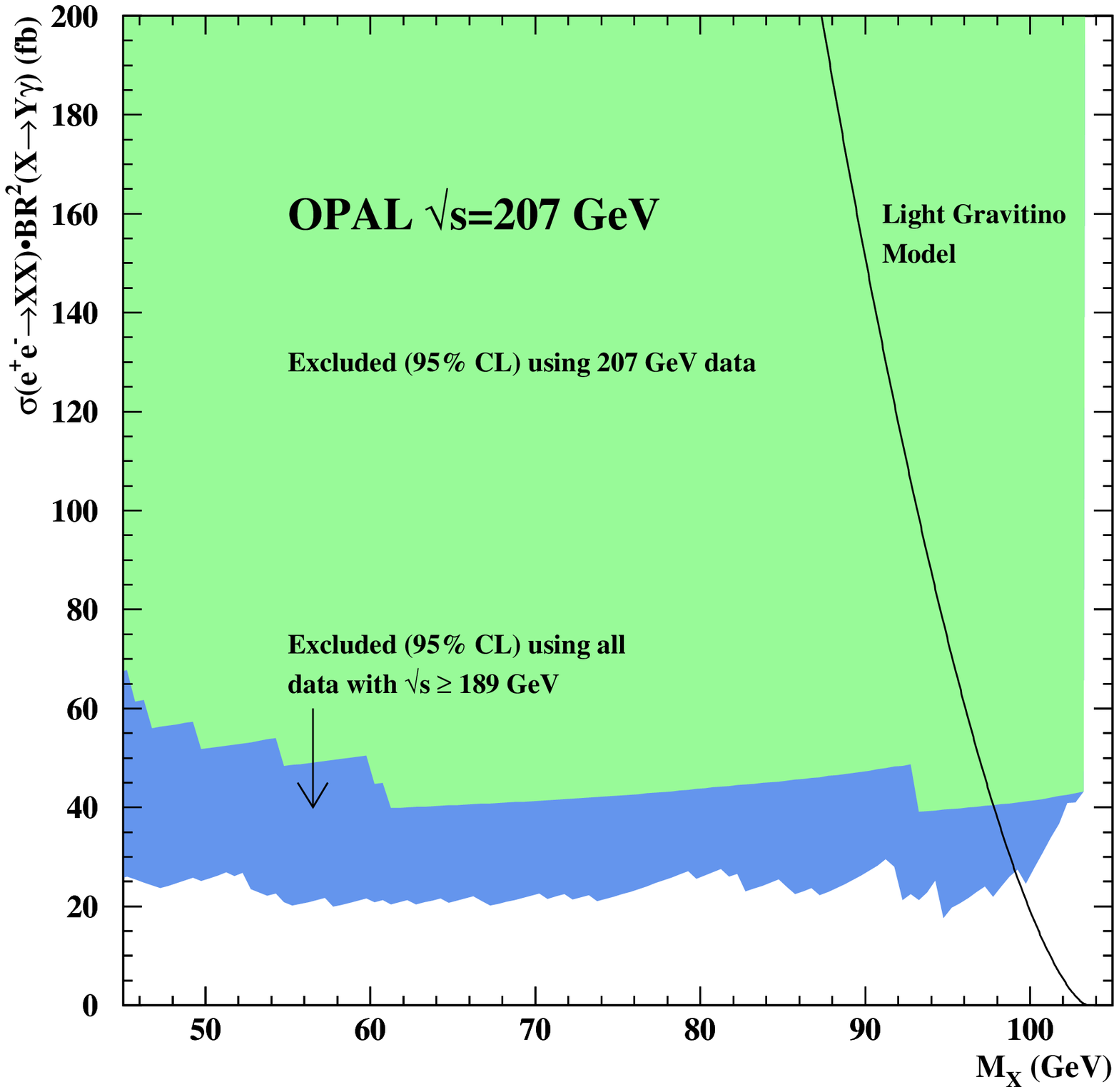}
\end{center}
\caption{
95\% CL upper limits on $\sigbrXX$ at 207 GeV for $\myzero$ 
obtained from all OPAL data with $\sqrt{s}\ge$ 189 GeV. The lightly 
shaded region shows the excluded region obtained using only the 
OPAL 207 GeV data sample. The darker region shows the exclusion
region obtained using all OPAL data with $\sqrt{s}\ge 189$~GeV, assuming 
that the cross-section scales as $\betax/s$. The line shows the prediction 
of an example light gravitino LSP model\cite{chang}. Within that model, 
$\lsp$ masses between 45 and 99~GeV are excluded at 95\% CL. These 
limits assume that particle X decays promptly.}
\label{my0_207_combined}
\end{figure}

Based on the efficiencies and the number of selected events, we calculate 
95\% CL upper limits on $\sigbrXX$ for $\myzero$ as a function of $\mx$, in
each region of centre-of-mass energy. 
The last two columns of Table~\ref{tab:limits} show the range of limits obtained
from each of the data samples, for $\mx$ values from 45 GeV up to the kinematic limit.
Figure~\ref{my0_207_combined} shows the limit obtained from the 207 GeV data sample, 
as well as the combined limit obtained from the entire data sample with 
$\sqrt{s}\ge 189$ GeV assuming that the cross-section scales as $\betax/s$. 
For the mass range of interest ($\mx > 45$ GeV) the model-independent limits range 
between 45 and 70 fb while the combined limits range between 20 and 45 GeV.
These limits
\footnote
{
In the $70-80$ GeV region the limits are actually slightly worse than those
along the $\my$=0 axis of Figure~\ref{mxmy_207_combined} despite the more efficient 
background suppression of the $\mxmax$ cut, relative to the kinematic consistency cuts
applied in the general case. This is due to a deficit of selected events in this region, 
compared to the expected background when using the general kinematic consistency 
requirements.
} 
can be used to set model-dependent 
limits on the mass of the lightest neutralino in supersymmetric models in which the NLSP is
the lightest neutralino and the LSP is a light gravitino 
($\PX=\lsp, \PY=\Gravitino$). Shown in Figure~\ref{my0_207_combined}, as a 
dotted line, is the (Born-level) cross-section prediction from a specific light 
gravitino LSP model\cite{chang} in which the neutralino composition is purely 
bino, with $m_{\tilde{e}_R} = 1.35 m_{\lsp}$ and 
$m_{\tilde{e}_L} = 2.7 m_{\lsp}$. Within the framework of this model, $\lsp$ 
masses between 45 and 99.0 GeV are excluded at 95\% CL. 

As described in section 2, the efficiencies over the full angular range have
been obtained using isotropic angular distributions for the production and decay of
$\PX$. The validity of this model  has been examined based on the angular 
distributions calculated for photino pair production in reference\cite{ELLHAG}. 
For models proposed in reference\cite{gravitinos}, the production angular 
distributions are more central and so this procedure is conservative. For a 
$1 + \cos^2{\theta}$ production angular distribution expected for t-channel 
exchange of a very heavy particle according to reference\cite{ELLHAG}, the 
relative efficiency reduction would be less than 2\% at all points in 
the ($\mx,\my$) plane.

\section{Conclusions}
We have searched for events with a final state consisting of two or three photons
and large missing energy, in data taken with the OPAL detector at LEP, at  
centre-of-mass energies in the range of $192-209$ GeV.

The selection requires at least two photons with scaled energy
$x_{\gamma}>0.05$ within the polar angle region $15^{\circ}<{\theta}<165^{\circ}$
or at least two photons with energy $E_{\gamma}>1.75$ GeV with one satisfying
$\acosthe < 0.8$ and the other satisfying $15^{\circ}<{\theta}<165^{\circ}$.
In each case, the requirement $p_T^{\gamma\gamma}/E_{\rm beam} > 0.05$
is also applied. There are 54 events selected. The KK2f prediction for the 
contribution from $\eetonngggbra$ is $57.2 \pm 1.3$ events; expected contributions 
from other sources sum  to $1.2\pm{0.3}$ events. The number of events observed in 
the data and their kinematic distributions are consistent with Standard Model 
expectations. Limits on new physics processes of the form
\mbox{$\rm \sigma ( {\mathrm e^+e^-\rightarrow XX} ) 
\cdot BR^2 ( {\mathrm X\rightarrow Y\gamma} )$}
are set separately at energies of 192, 196, 200, 202, 205 
and 207 GeV. In addition, combined limits are set at $\roots = 207$ GeV, assuming 
a $\betax/s$ scaling of the production cross-section 
\mbox{$\rm \sigma ( {\mathrm e^+e^-\rightarrow XX} )$}. 
From the full OPAL data sample with $\sqrt{s}\ge 189$ GeV,
we derive 95\% CL upper limits on $\sigbrXX$ ranging from 10 to 
60~fb for the general case of massive $\PX$ and $\PY$. 
For the special case of $\myzero$, the 95\% CL upper limits on $\sigbrXX$ range 
from 20 to 45 fb, for $\mx > 45$ GeV. These results are used to place 
model-dependent lower limits on the $\lsp$ mass in a specific light gravitino 
LSP model\cite{chang}. Masses between 45 and 99 GeV are excluded at 95\% CL. 
All limits assume that particle $\PX$ decays promptly.

\medskip
\bigskip\bigskip\bigskip
\appendix
\par
\section*{Acknowledgements}
\par
We particularly wish to thank the SL Division for the efficient operation
of the LEP accelerator at all energies
 and for their close cooperation with
our experimental group.  In addition to the support staff at our own
institutions we are pleased to acknowledge the  \\
Department of Energy, USA, \\
National Science Foundation, USA, \\
Particle Physics and Astronomy Research Council, UK, \\
Natural Sciences and Engineering Research Council, Canada, \\
Israel Science Foundation, administered by the Israel
Academy of Science and Humanities, \\
Benoziyo Center for High Energy Physics,\\
Japanese Ministry of Education, Culture, Sports, Science and
Technology (MEXT) and a grant under the MEXT International
Science Research Program,\\
Japanese Society for the Promotion of Science (JSPS),\\
German Israeli Bi-national Science Foundation (GIF), \\
Bundesministerium f\"ur Bildung und Forschung, Germany, \\
National Research Council of Canada, \\
Hungarian Foundation for Scientific Research, OTKA T-038240, 
and T-042864,\\
The NWO/NATO Fund for Scientific Research, the Netherlands.\\


\newpage
%
\begin{table}
\centering
\begin{tabular}{|c|c|c|c|r|r|c|c|} \hline
Sample & $\cal{L}$ (pb$^{-1}$) & $\sqrt{s}$ (GeV) & $<\sqrt{s}>$ & $N_{obs}$ & $N^{\nngggbra}_{exp}\phantom{00}$ & $\mathrm {\sigma}_{meas}^{\nngggbra}$(pb) & $\mathrm {\sigma}_{KK2f}^{\nngggbra}$(pb) \\ \hline \hline
192 &  28.9 & $190-194$ & 191.6 &        4  & $ 4.26\pm{0.11}$     & $0.21\pm{0.10}$ & $0.222\pm{0.003}$ \\ \hline
196 &  72.3 & $194-198$ & 195.6 &        5  & $ 9.97\pm{0.25}$     & $0.11\pm{0.05}$ & $0.215\pm{0.002}$ \\ \hline
200 &  74.8 & $198-201$ & 199.5 &       14  & $10.10\pm{0.25}$     & $0.29\pm{0.08}$ & $0.207\pm{0.001}$ \\ \hline
202 &  39.2 & $201-203$ & 201.7 &  6  & $ 5.21\pm{0.14}$     & $0.23\pm{0.10}$ & $0.203\pm{0.002}$ \\ \hline 
205 &  79.1 & $203-206$ & 205.0 & 10  & $10.34\pm{0.26}$     & $0.19\pm{0.06}$ & $0.198\pm{0.001}$ \\ \hline
207 & 132.2 & $206-209$ & 206.6 & 15  & $17.28\pm{0.43}$     & $0.17\pm{0.04}$ & $0.196\pm{0.001}$ \\ \hline
\end{tabular}
\caption{Results of the selection applied to the OPAL 1999 and 2000 data
samples. Shown for each sub-sample are the integrated luminosity $\cal{L}$, the centre-of-mass
energy range, the luminosity-weighted mean centre-of-mass energy, the numbers of events observed 
and expected, and the measured and predicted cross-section for the process $\eetonngggbra$, 
within the kinematic acceptance of the selection. Predicted values were obtained using the KK2f 
Monte Carlo generator. The errors shown are the sum of the statistical and systematic uncertainties.
}
\label{tab:g2_xsec_new}
\end{table}

%
%
\begin{table}
\centering
\begin{tabular}{|c||c|c|c|c|}
\hline
$\mx$ (GeV) & $\my$=0        & $\my=\mx /2$   & $\my=\mx-10$   & $\my=\mx-5$   \\ \hline \hline
102.5 & $74.5\pm{1.2}$ & $74.7\pm{1.1}$ & $63.2\pm{1.3}$ & $33.8\pm{1.5}$   \\ \hline
100   & $74.5\pm{1.2}$ & $74.4\pm{1.1}$ & $61.4\pm{1.3}$ & $32.3\pm{1.5}$     \\ \hline
90    & $74.3\pm{1.2}$ & $75.1\pm{1.1}$ & $60.4\pm{1.4}$ & $36.2\pm{1.5}$      \\ \hline
80    & $73.2\pm{1.2}$ & $73.5\pm{1.2}$ & $65.4\pm{1.3}$ & $37.8\pm{1.5}$      \\ \hline
70    & $74.1\pm{1.2}$ & $71.7\pm{1.2}$ & $62.0\pm{1.4}$ & $39.0\pm{1.5}$      \\ \hline
60    & $73.8\pm{1.1}$ & $71.5\pm{1.2}$ & $62.5\pm{1.4}$ & $41.2\pm{1.5}$      \\ \hline
50    & $72.1\pm{1.2}$ & $71.5\pm{1.2}$ & $65.2\pm{1.3}$ & $43.5\pm{1.5}$      \\ \hline
\end{tabular}
\caption[]{Selection efficiencies (\%) 
for the process $\eetoXX$, $\XtoYg$ at $\roots = 206$~GeV for various
$\mx$ and $\my$ (GeV), after application of kinematic-consistency cuts.
Not shown are the values for $\my = 20$~GeV, $\my = \mx-15$~GeV and $\my = \mx-2.5$~GeV. 
The errors shown are due to Monte Carlo statistics only.
}
\label{tab:g2_eff_206}
\end{table}

\begin{table}
\centering
\begin{tabular}{|c||r|r|r|r|}
\hline
$\sqrt{s}$ & $\sigma_{95}^{min}(\mx,\my)$ & $\sigma_{95}^{max}(\mx,\my)$ & $\sigma_{95}^{min}(\mx)$ & $\sigma_{95}^{max}(\mx)$ \\ \hline  \hline
192 & 138 fb  & 296 fb & 143 fb & 288 fb  \\ \hline 
196 &  60 fb  & 125 fb &  71 fb &  87 fb \\ \hline
200 &  57 fb  & 278 fb &  57 fb & 237 fb \\ \hline
202 & 105 fb  & 323 fb & 106 fb & 206 fb \\ \hline
205 &  52 fb  & 183 fb &  70 fb & 130 fb \\ \hline
207 &  31 fb  &  90 fb &  45 fb &  70 fb \\ \hline
\end{tabular}
\caption[]{
Results of individual limit calculations at each centre-of-mass energy. The first 
column shows the data sample. The second and third columns show the maximum and 
minimum 95\% CL limits on $\sigbrXX$  in the $(\mx, \my)$ plane, for the case of 
massive $\PY$ for $\mx > M_{\mathrm Z}/2$ and $\mx-\my < 5$ GeV. The last two columns show
the minimum and maximum 95\% CL limits obtained for the special case of $\my\approx$0,
for $\mx$ values between 45 GeV and the kinematic limit.
}
\label{tab:limits}
\end{table}

\begin{table}
\centering
\begin{tabular}{|c||c|c|c|c|}
\hline
$\mx$ & Selection & Selection efficiency (\%) with & $\rm N_{data}$ & 
$\rm N_{\nngggbra}$ \\
$(GeV)$ & efficiency (\%) & $\mxmax>\mx-5$ GeV &  & \\ \hline
\hline
102.5 & $75.6 \pm 1.1$  & $73.6 \pm 1.3$ & 2 & $1.28\pm{0.08}$ \\ \hline
100   & $75.7 \pm 1.1$  & $72.7 \pm 1.3$ & 2 & $2.08\pm{0.10}$ \\ \hline
90    & $74.9 \pm 1.1$  & $72.5 \pm 1.2$ & 3 & $4.14\pm{0.16}$ \\ \hline
80    & $73.7 \pm 1.2$  & $71.3 \pm 1.2$ & 4 & $6.13\pm{0.22}$ \\ \hline
70    & $74.5 \pm 1.2$  & $71.7 \pm 1.2$ & 5 & $8.51\pm{0.28}$ \\ \hline
60    & $73.9 \pm 1.2$  & $72.2 \pm 1.2$ & 5 & $11.25\pm{0.34}$ \\ \hline
50    & $72.3 \pm 1.2$  & $69.5 \pm 1.2$ & 10 & $14.85\pm{0.42}$ \\ \hline
\end{tabular}
\caption[]{
Selection efficiencies as a function of $\mx$  for the process $\eetoXX$, 
$\XtoYg$, for $\myzero$ at $\roots = 206$ GeV. The second column shows the 
efficiency of the general selection. The third column shows the efficiency 
including the additional cut on $\mxmax$.  The errors on the efficiencies are
statistical only. The fourth column
shows the number of events from the $205-207$ GeV data sample consistent with 
the mass value $\mx$. The last column shows the corresponding number of expected 
events from the process $\eetonngggbra$, obtained using KK2f, along with the 
corresponding uncertainty (statistical plus systematic). 
}
\label{tab:g2_eff_my0_206}
\end{table}
%
%
%
\clearpage
\newpage
\newpage
\newpage
\newpage

\begin{thebibliography}{99}
\bibitem{OPALSP189}
  OPAL Collab., G. Abbiendi et al.,
  Eur. Phys. J. {\bf C18} (2000) 253.
\bibitem{OPALSP183}
  OPAL Collab., G. Abbiendi et al.,
  Eur. Phys. J. {\bf C8} (1999) 23.  
\bibitem{OPALSP172}
  OPAL Collab., K. Ackerstaff et al.,
  Eur. Phys. J. {\bf C2} (1998) 607.
\bibitem{LEP2AP}
  ALEPH Collab., R. Barate et al.,
  Eur. Phys. J. {\bf C28} (2003) 1.;\newline
  DELPHI Collab.,  P. Abreu et al.,
  CERN-EP-2003-093. Submitted to Eur. Phys. J.;\newline
  L3 Collab., J. Abdallah et al.,
  Phys. Lett. {\bf B587} (2004) 16.  
\bibitem{chang}
C.Y. Chang and G.A. Snow, UMD/PP/97-57;\newline
K. S. Babu, C. Kolda and F. Wilczek, Phys. Rev. Lett. {\bf 77} (1996) 3070.
%
\bibitem{OPAL_Hgg209}
  OPAL Collab., G. Abbiendi et al.,
  Phys. Lett. {\bf B544} (2002) 44.
\bibitem{OPALQGC}
  OPAL Collab., G. Abbiendi et al.,
  CERN-EP-2004-003. Submitted to Phys. Rev. D
\bibitem{OPAL-detector}
  \OPALColl, K.~Ahmet et~al., \NIM\ {\bf A305} (1991) 275;\newline
  S.~Anderson et~al., \NIM\  {\bf A403} (1998) 326;\newline
  B.E.~Anderson et~al., IEEE Transactions on Nuclear Science {\bf 41}
  (1994) 845.
\bibitem{NUNUGPV98}G.~Montagna, M.~Moretti, O.~Nicrosini and 
F.~Piccinini, Nucl. Phys. {\bf B541} (1999) 31.
\bibitem{KK2f} 
S. Jadach, B.F.L. Ward and Z. W\c{a}s, Phys. Lett. {\bf B449} (1999) 97; \\
S. Jadach, B.F.L. Ward and Z. W\c{a}s, Comp. Phys. Comm. 130 (2000) 260. 
\bibitem{RADCOR}
  F.A.~Berends and R.~Kleiss,
  Nucl.\ Phys.\ {\bf B186} (1981) 22.   
\bibitem{BHWIDE}
S.~Jadach, W.~Placzek and B.F.L.~Ward, 
Phys. Lett. {\bf B390} (1997) 298. 
\bibitem{TEEGG}
  D.~Karlen, Nucl.\ Phys.\ {\bf B289} (1987) 23.
\bibitem{KORALW}
  S. Jadach et al., Comp. Phys. Comm. {\bf 119} (1999) 272.
\bibitem{GRC4F}
  J.~Fujimoto et al., Comp. Phys. Comm. {\bf 100} (1997) 128.
%
\bibitem{KORALZ}
S.~Jadach, B.F.L.~Ward and Z.~W\c{a}s, Comp. Phys. Comm. {\bf 79} (1994) 503.\newline
Version 4.02 was used including a recommended correction to the 
  NDIST0 subroutine.
%
\bibitem{BDK}
 F.A.~Berends, P.H.~Daverveldt and R.~Kleiss, \NPhys\ {\bf B253} (1985) 421; \\
 F.A.~Berends, P.H.~Daverveldt and R.~Kleiss, \CPC\ {\bf 40} (1986) 271, 285
 and 309.
%
%
\bibitem{VERMASEREN}
J.~A.~M.~Vermaseren, Nucl. Phys. {\bf B229} (1983) 347.
\bibitem{SUSYGEN}
S. Katsanevas and S. Melachronios, in Physics at LEP2, 
edited by G. Alterelli, T.~Sj\"ostrand and F. Zwirner, CERN/96-01, Vol.2 (1996) 328. 
%
\bibitem{GOPAL}
  J.~Allison et~al., \NIM\  {\bf A317} (1992) 47.
%
\bibitem{LEP1XX}
  OPAL Collab., M.Z.~Akrawy et al.,
  Phys. Lett. {\bf B248} (1990) 211;\newline
  ALEPH Collab., D.~Decamp et al., 
  Phys. Rep. {\bf 216} (1992) 253;\newline
  L3 Collab., M.~Acciarri et al.,
  Phys. Lett. {\bf B350} (1995) 109.
\bibitem{systerr} R.D.~Cousins and V.L.~Highland, 
  \NIM\ {\bf A320} (1992) 331.
\bibitem{gravitinos2}
J.L. Lopez and D.V. Nanopoulos, Mod. Phys. Lett. {\bf A11} (1996) 2473;\newline
Phys. Rev. {\bf D55} (1997) 4450.
\bibitem{ELLHAG}
J. Ellis and J.S. Hagelin, Phys. Lett. {\bf B122} (1983) 303.
\bibitem{gravitinos}
S. Dimopoulos et al., Phys. Rev. Lett. {\bf 76} (1996) 3494;\newline
D.R. Stump, M. Wiest, C.P. Yuan, Phys. Rev. {\bf D54} (1996) 1936;\newline
S. Ambrosanio et al., Phys. Rev. {\bf D54} (1996) 5395.
\end{thebibliography}
\end{document}